\pdfoutput=1
\documentclass[10pt,aps,twocolumn,prl,superscriptaddress]{revtex4-2}
\usepackage{graphicx}
\usepackage{braket}
\usepackage[pagewise]{lineno}
\usepackage[breaklinks=true,colorlinks,citecolor=blue,linkcolor=blue,urlcolor=blue]{hyperref}
\usepackage{tabularx}
\usepackage{xcolor}
\usepackage{amsmath}
\usepackage{amssymb,braket}
\usepackage{mathrsfs}
\usepackage{multirow}
\usepackage{bm}

\usepackage{ulem}
\usepackage{float}
\usepackage{appendix}
\usepackage{changes}
\usepackage{url}
\usepackage{siunitx}
\usepackage{graphicx}

\usepackage{lineno} 
\usepackage[normalem]{ulem}

\begin{document}

\title{Strain-engineered nanoscale spin polarization reversal in diamond nitrogen-vacancy centers}

\author{Zhixian Liu}
\thanks{These authors contributed equally to this work.}
\affiliation{Laboratory of Spin Magnetic Resonance, School of Physical Sciences, Anhui Province Key Laboratory of Scientific Instrument Development and Application, University of Science and Technology of China, Hefei 230026, China}

\author{Jiahao Sun}
\thanks{These authors contributed equally to this work.}
\affiliation{Laboratory of Spin Magnetic Resonance, School of Physical Sciences, Anhui Province Key Laboratory of Scientific Instrument Development and Application, University of Science and Technology of China, Hefei 230026, China}

\author{Ganyu Xu}
\thanks{These authors contributed equally to this work.}
\affiliation{Laboratory of Spin Magnetic Resonance, School of Physical Sciences, Anhui Province Key Laboratory of Scientific Instrument Development and Application, University of Science and Technology of China, Hefei 230026, China}

\author{Bo Yang}
\affiliation{Laboratory of Spin Magnetic Resonance, School of Physical Sciences, Anhui Province Key Laboratory of Scientific Instrument Development and Application, University of Science and Technology of China, Hefei 230026, China}
\affiliation{Hefei National Research Center for Physical Sciences at the Microscale, Hefei 230026, China}

\author{Yuhang Guo}
\affiliation{Laboratory of Spin Magnetic Resonance, School of Physical Sciences, Anhui Province Key Laboratory of Scientific Instrument Development and Application, University of Science and Technology of China, Hefei 230026, China}
\affiliation{Hefei National Research Center for Physical Sciences at the Microscale, Hefei 230026, China}

\author{Yu Wang}
\affiliation{Institute of Geosciences, Goethe University Frankfurt, Frankfurt 60438, Germany}

\author{Cunliang Xin}
\affiliation{Laboratory of Spin Magnetic Resonance, School of Physical Sciences, Anhui Province Key Laboratory of Scientific Instrument Development and Application, University of Science and Technology of China, Hefei 230026, China}
\affiliation{Hefei National Research Center for Physical Sciences at the Microscale, Hefei 230026, China}

\author{Hongfang Zuo}
\affiliation{Laboratory of Spin Magnetic Resonance, School of Physical Sciences, Anhui Province Key Laboratory of Scientific Instrument Development and Application, University of Science and Technology of China, Hefei 230026, China}
\affiliation{Hefei National Research Center for Physical Sciences at the Microscale, Hefei 230026, China}

\author{Mengqi Wang}
\email{mqw@ustc.edu.cn}
\affiliation{Laboratory of Spin Magnetic Resonance, School of Physical Sciences, Anhui Province Key Laboratory of Scientific Instrument Development and Application, University of Science and Technology of China, Hefei 230026, China}
\affiliation{Hefei National Research Center for Physical Sciences at the Microscale, Hefei 230026, China}

\author{Ya Wang}
\email{ywustc@ustc.edu.cn}
\affiliation{Laboratory of Spin Magnetic Resonance, School of Physical Sciences, Anhui Province Key Laboratory of Scientific Instrument Development and Application, University of Science and Technology of China, Hefei 230026, China}
\affiliation{Hefei National Research Center for Physical Sciences at the Microscale, Hefei 230026, China}
\affiliation{Hefei National Laboratory, University of Science and Technology of China, Hefei 230088, China}

\date{\today}

\begin{abstract}
The ability to control solid-state quantum emitters is fundamental to advancing quantum technologies.
The performance of these systems is fundamentally governed by their spin-dependent photodynamics, yet conventional control methods using cavities offer limited access to key non-radiative processes. 
Here we demonstrate that anisotropic lattice strain serves as a powerful tool for manipulating spin dynamics in solid-state systems. Under high pressure, giant shear strain gradients trigger a complete reversal of the intrinsic spin polarization, redirecting ground-state population from $\ket{0}$ to $\ket{\pm 1}$ manifold. We show that this reprogramming arises from strain-induced mixing of the NV center’s excited states and dramatic alteration of intersystem crossing, which we quantify through a combination of opto-magnetic spectroscopy and a theoretical model that disentangles symmetry-preserving and symmetry-breaking strain contributions.
Furthermore, the polarization reversal is spatially mapped with a transition region below 120 nm, illustrating sub-diffraction-limit control. Our work establishes strain engineering as a powerful tool for tailoring quantum emitter properties, opening avenues for programmable quantum light sources, high-density spin-based memory, and hybrid quantum photonic devices.

\end{abstract}

\maketitle


Solid-state quantum emitters, with their intrinsic spin degrees of freedom, are leading candidates for developing advanced quantum technologies\cite{du2024single,pompili2021realization,knaut2024entanglement,hermans2022qubit,wan2020large,bradley2019ten}. The performance of these systems is dictated by the emitter's radiative lifetime and the non-radiative intersystem crossing (ISC) rate, which constitute the physical basis for efficient spin initialization and readout\cite{goldman2015phonon,goldman2015state,thiering2018theory,liu2024silicon}. For decades, the primary strategy for controlling these properties has relied on integrating emitters into passive photonic nanostructures\cite{albrecht2013coupling,grange2017reducing,rugar2021quantum,ding2024high}, such as cavities and waveguides, to engineer the local density of optical states via the Purcell effect. While successful in accelerating photon emission rates, this approach primarily influences the radiative pathway, leaving non-radiative processes and the intricate electronic structure governing spin dynamics largely unaltered. Here, We introduce a novel strain-engineering strategy that offers direct, in situ, and tailored control over a quantum emitter's structural symmetry and electronic properties. This approach enables direct and simultaneous manipulation of both its optical lifetime and spin-dependent photodynamics.

We demonstrate this new scheme using nitrogen-vacancy (NV) centers in diamond, a leading platform for quantum photonics and nanoscale quantum sensing with diverse applications from biology to materials science \cite{balasubramanian2008nanoscale,maze2008nanoscale,zhao2012sensing,staudacher2013nuclear,rugar2015proton,shi2015single,sushkov2014all,kucsko2013nanometre, degen2017quantum,ji2024correlated, wang2024imaging}.
In particular, their unique compatibility with diamond anvil cell (DAC) technology has recently enabled quantum measurements under extreme pressures exceeding 130 GPa \cite{bhattacharyya2024imaging,wang2024imaging,hilberer2023enabling} and applications in characterizing high-temperature superconductivity Messiner effect and pressure-driven magnetic phase transition \cite{bhattacharyya2024imaging,wang2024imaging,hilberer2023enabling,yip2019measuring,lesik2019magnetic,hsieh2019imaging}.  
However, a critical knowledge gap persists regarding their response to complex strain environments, where the $C_{3v}$ symmetry of NV center is broken by specific components of the non-hydrostatic stress tensor, leading to strongly altered spin-optical dynamics\cite{doherty2014electronic, hilberer2023enabling, dai2022optically}. 
Understanding the strain coupling mechanisms could help us develop next-generation sensors for higher pressure regimes and harness strain as a powerful tool to program quantum emitters beyond the reach of conventional methods.

 \begin{figure*}[ht]
    \centering
    \includegraphics[width=1.8\columnwidth]{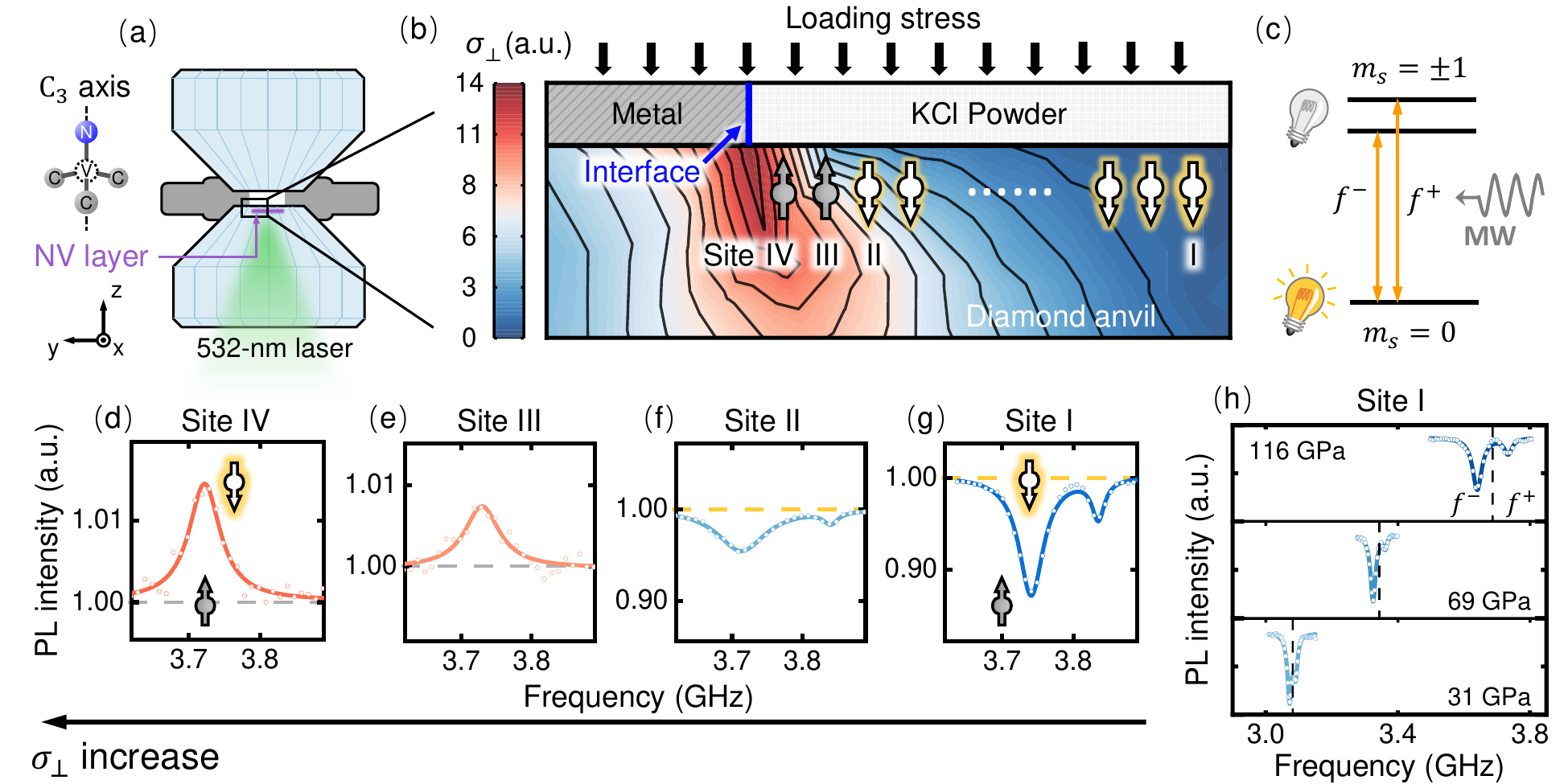}
    \caption{Strain-dependent spin polarization in NV centers under high pressure. (a) Schematic cross-section of the diamond anvil cell. 
    The laboratory coordinate frame is aligned with the NV symmetry axes: $z\parallel[111],  x\parallel[\bar{1}10], y\parallel[\bar{1}\bar{1}2]$. (b) Finite-element simulation of the symmetry-breaking stress value $\sigma_{\perp}$ distribution in the anvil (SI.I(c)). The grey and white spin arrows denote the corresponding local spin polarization of NV centers, indicating the transition from the conventional down-polarized state ($\ket{m_s=0}$) to the reversed up-polarized state ($\ket{m_s=\pm 1}$). (c) Energy level diagram of the NV center ground-state triplet. The $\ket{m_s=0}$ state exhibits bright photoluminescence (PL), while the $\ket{m_s=\pm 1}$ states are dark. These spin states can be coherently manipulated using microwaves (MW) via electron spin resonance. (d)-(g) Optically detected magnetic resonance (ODMR) spectra from Site I to Site IV under approximately 129 GPa. The color gradient (blue to red) tracks the transition from negative to positive ODMR contrast, correlating with increasing symmetry-breaking stress. (h) ODMR spectra from Site I under increasing pressure (31 - 116 GPa). The dashed lines mark the blue shift of the $\ket{0}\rightarrow\ket{+1}$ ($ f^+$) and $\ket{0}\rightarrow\ket{-1}$ resonance frequencies ($ f^-$). 
  }\label{fig1}
  \end{figure*}
  
In this work, we bridge this gap by revealing the fundamental mechanisms of strain-spin coupling in NV centers. With this insight, we first demonstrate tailored control over the spin polarization dynamics, and subsequently achieve nanoscale, deterministic reversal of the optical spin polarization via engineered strain gradients. Our results establish strain as a powerful and versatile approach  to control the properties of quantum emitters, introducing a new dimension to regulate spin-optical interactions. The method developed here can also be readily implemented in various nanoscale sensors, such as spin defects in silicon carbide \cite{wang2023magnetic} and hexagonal boron nitride \cite{gottscholl2021spin}.

Our experiments are performed in a (111)-oriented diamond anvil cell incorporating a near-surface ensemble of NV centers, whose $C_{3v}$ symmetry axes are aligned perpendicular to the anvil surface (SI.I(b)).
At the cell edge, we utilize the interface between metallic rhenium and KCl powder---materials with highly mismatched elastic moduli---to generate a controlled shear strain gradient inside the diamond (Fig.1(a, b)).
(1) In regions far from this metal–powder interface (Site I, near the chamber center), the stress state is well described by an idealized semi-infinite slab model\cite{ruoff1991closing}, where the total stress decomposes into a hydrostatic component and a uniaxial component perpendicular to the anvil surface. For the implanted NV centers, this uniaxial stress is aligned with the $C_{3v}$ axis, thus constituting a symmetry-preserving perturbation ${\bm{\sigma}_{\rm{C_{3v}}}}$(End Matter).
(2) In contrast, near the metal–powder interface (Site IV), the complex stress field substantially deviates from the idealization and contains prominent symmetry-breaking contributions. These are represented by the term ${\bm{\sigma}_{\rm{non-C_{3v}}}}$ (End Matter), which significantly alters the NV center’s spin Hamiltonian and spin-optical properties. Spatial mapping across the strain gradient from Site I to Site IV provides a controlled platform to decouple the competing effects of symmetry-preserving and symmetry-breaking strain, allowing us to isolate and quantitatively compare their distinct influences on the NV properties.

We begin by probing the spin states of NV centers under varying strain conditions using optically detected magnetic resonance (ODMR) measurements at zero magnetic field. These measurements reveal fundamentally distinct behaviors in the optical spin initialization process, governed by the local strain symmetry. In symmetry-preserving regions (Sites I and II), we observe conventional ODMR spectra characterized by negative dips (Fig. \ref{fig1}(f, g)), accompanied by a pressure-induced blue shift of the resonance frequency from 2.87 GHz at ambient pressure to 3.79 GHz ($\sim$129 GPa) (Fig. \ref{fig1}(g, h)), consistent with previous reports under quasi-hydrostatic  conditions\cite{wang2024imaging}. In contrast, as we probe NV centers located near high-shear-strain regions (Sites III and IV), the ODMR response undergoes a complete inversion: the characteristic negative dips transition into positive features (Fig. \ref{fig1}(d) and (e)). This signal inversion provides direct evidence of a strain-governed switching of the optical spin initialization pathway. NV centers in symmetry-preserving regions follow the conventional route, initializing into the bright $\ket{0}$ state and yielding negative ODMR contrasts ((Fig. \ref{fig1}(c))). By comparison, in regions where symmetry-breaking strain is significant, the initialization preferentially populates the dark $\ket{\pm 1}$ states, resulting in positive ODMR signals. 

\begin{figure}[htbp]
  \includegraphics[width=0.92\linewidth]{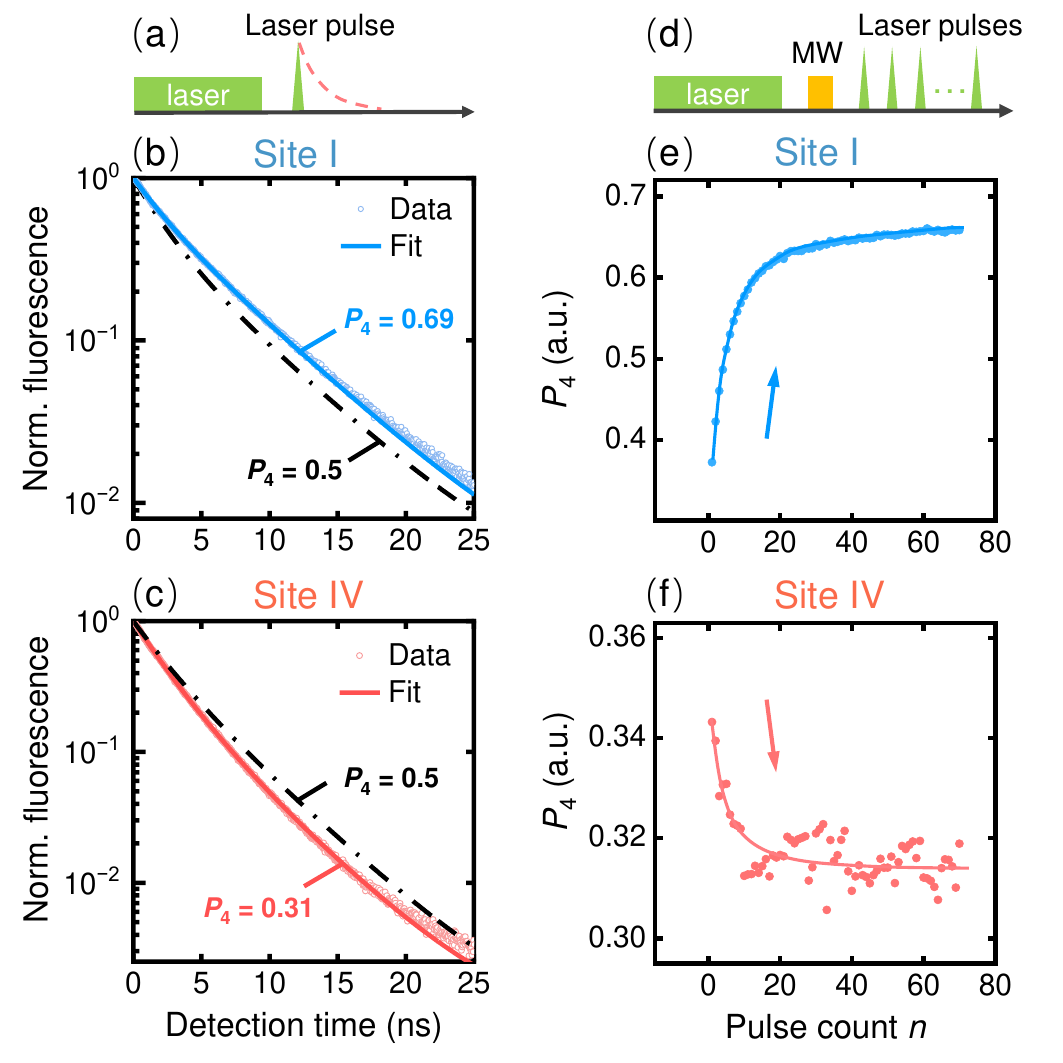}
  \caption{\label{fig2} Quantifying excited-state spin polarization dynamics. 
  (a) Schematic of the time-resolved fluorescence measurement sequence. A picosecond-laser pulse excitation (green triangle) probes the excited-state spin population following continuous-wave laser initialization (green rectangle). 
  (b-c) Fluorescence decay curve for NV centers at Site I (blue curve) and Site IV (red curve). Solid lines represent tri-exponential fits to the data. A reference decay curve for $P_4$ = 0.5 (black dashed-dotted line) is included for comparison.
   (d) The multi-pulse polarization measurement Following optical initialization and a polarization-inverting microwave $\pi$ pulse (yellow rectangle), a train of picosecond laser pulses monitors the evolution of the spin polarization. 
   (e-f) Evolution of bright state population $P_4$ as a function of the ps-laser pulse number for Site I (blue) and Site IV (red). Solid lines serve as guidelines highlighting the opposing polarization trends between the two sites.
  }
  \end{figure}

The reversed spin initialization is further confirmed by time-resolved fluorescence measurements under picosecond (ps) pulsed laser excitation (Fig. \ref{fig2}(a)).
The fluorescence intensity decay follows a multi-exponential form, $I(t) \propto \sum_{j=4}^{6} P_j e^{-t/\tau_j}$, with the normalization condition $\sum_{j=4}^{6} P_{j} = 1$, where $P_j$ and $\tau_j$ represent the population and lifetime of the excited state $\ket{j}(j =4,5,6)$. The overall polarization of the excited state is quantified by $P_4$, the population of the bright state $\ket{4}$, which exhibits the longest lifetime among the three states $(\tau_4>\tau_5,\tau_6)$ (SI.II(a,b)). 
By fitting the fluorescence decay curves (Fig. \ref{fig2}(b, c)), we extract the polarization $P_4$ for Sites I and IV. To ensure fitting reliability and minimize the number of free parameters, the lifetimes $\tau_j$ are predetermined by preparing a thermalized spin state via spin relaxation (Fig. S5, SI.II(b)). Remarkably,  the extracted $P_4$ values reveal a striking contrast between the two strain regimes: the spin polarization reverses from $P_4 = 0.69$ (Site I) to $P_4 = 0.31$ (Site IV), corresponding to a preferential initialization into the bright and dark states, respectively. 
This polarization reversal is further demonstrated by the opposing dynamical behaviors observed from $\pi$-reversed initial states, which evolve towards distinct steady-state polarizations under multi-pulse excitation (Fig. \ref{fig2}(d-f)).

\begin{figure*}[htbp]
  \includegraphics[width=0.9\linewidth]{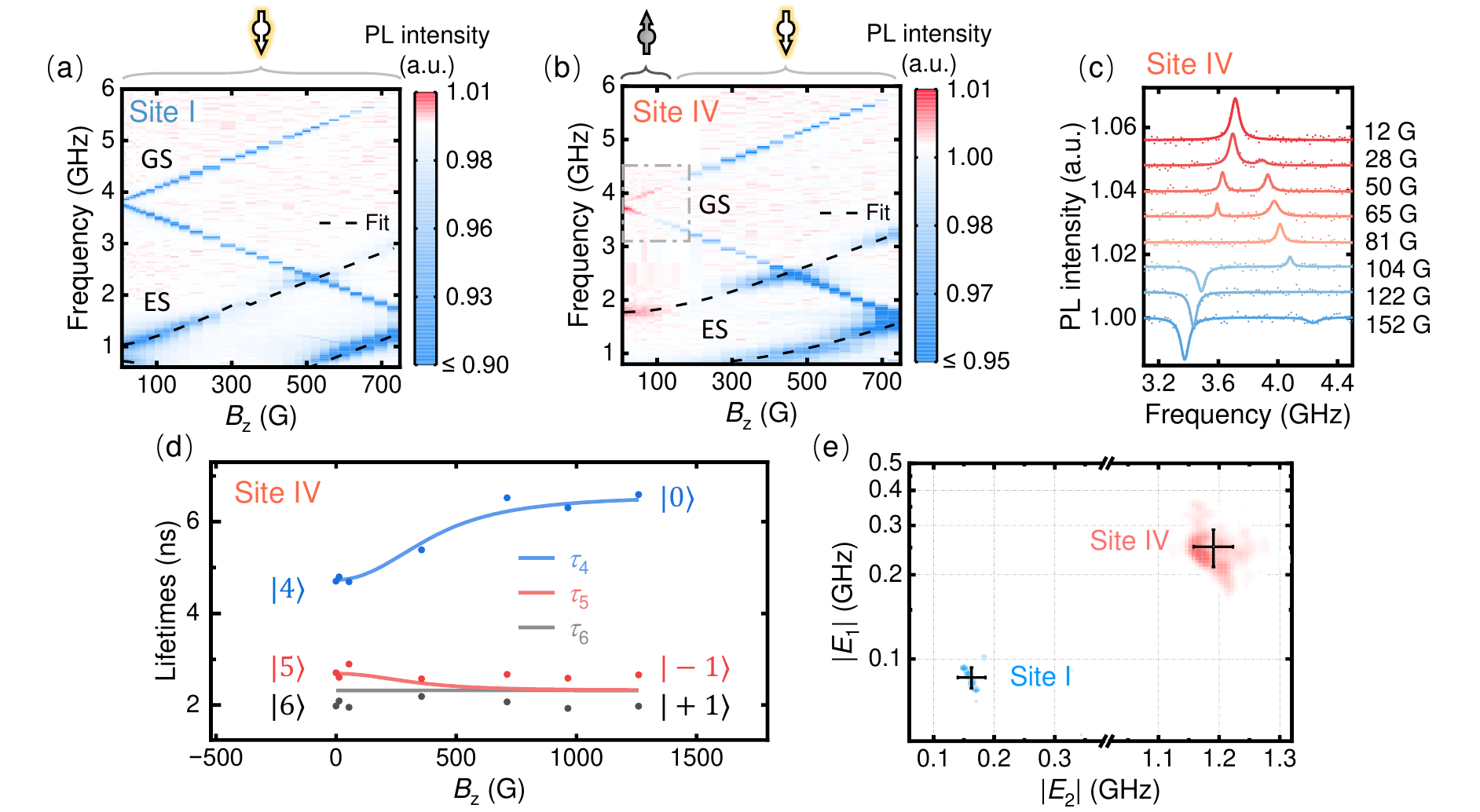}
  \caption{\label{fig3} Magnetic field control of strain-induced spin reversal. 
  (a-b) Magnetic field dependence of the ODMR spectrum of NV centers at Site I and Site IV, with $B_z$ applied along the NV symmetry axis. The red and blue colors in the heatmap denote positive and negative ODMR contrasts, respectively. Black dashed lines represent eigenfrequency fits based on the excited-state (ES) Hamiltonian (Eq. 1). Unfitted spectra correspond to resonant transitions of the NV ground state (GS). 
  (c) ODMR spectra at Site IV under a magnetic field from 12 G to 152 G, corresponding to the region outlined by the grey dashed-dotted rectangle in (b).
  (d) Excited-state lifetimes $\tau_j$ at Site IV versus $B_z$. Increasing $B_z$ quenches the strain-induced superposition of excited states (from $\ket{j}(j=4,5,6)$ to $\ket{0},\ket{\pm 1}$), resulting in the observed lifetime variations. Data points represent experimental measurements. Solid curves represent the corresponding model fits in (e).
  (e) The range of $|E_1|$ and $|E_2|$ terms derives from joint fitting, quantifying the extent of the local symmetry-breaking strain at different sites. The error bars represent the standard errors of the fitting results.
  }
  \end{figure*}

The application of an axial magnetic field enables active and deterministic control over strain-induced spin polarization reversal. As shown in Fig. \ref{fig3}(a-c), the ODMR contrast at Site IV transitions from positive at low fields to negative at higher fields, whereas Site I remains unaffected. This result indicates that spin polarization reversal can be suppressed by an external magnetic field.
Fluorescence lifetime measurements (Fig. \ref{fig3}(d) and Fig. S5) reveal the underlying mechanism: By quenching the strain-induced mixing of excited states, the magnetic field recovers the well-defined spin eigenstates ($\ket{0}$ and $\ket{\pm 1}$), which alters their selective coupling to the metastable singlet state. Consequently, the transition rates are modified, as directly observed in the field-dependent variation of the excited-state lifetimes. This result unambiguously confirms the magnetic tuning of the optical spin initialization pathway.

Quantitatively understanding the effects of strain on excited states requires a systematic characterization of symmetry-breaking strain. Here, we provide a novel framework for strain analysis and unravel its impact on spin-optical dynamics. While strain-induced mixing between $\ket{0}$ and $\ket{\pm 1} $ is often neglected in prior studies \cite{barson2017nanomechanical,hsieh2019imaging,wang2024imaging,bhattacharyya2024imaging}, this coupling becomes essential for modeling excited states under large in-plane distortions. Our model accounts for excited-state coupling to complex strain environments with large symmetry-breaking strain via the Hamiltonian (\cite{udvarhelyi2018spin}, SI.II(c)):
\begin{align}\label{eq1}
H = 
\begin{bmatrix}
  D+\gamma_e B_z & E_1 & E_2 \\
  E_1^* & 0 & -E_1 \\
  E_2^* & -E_1^* & D- \gamma_e B_z
\end{bmatrix}
\end{align},
where $B_z$ denotes axial magnetic field and $\gamma_e$ denotes gyromagnetic ratio. $D$, $E_1$ and $E_2$ represent spin-strain coupling terms. Crucially, symmetry-preserving strain modulates the diagonal term $D$, leading to a blueshift of the zero-field splitting. In contrast, symmetry-breaking strain activates the off-diagonal couplings $E_1$ and $E_2$ (see End Matter).

At zero field ($B_z \sim 0$ G), the $E_2$ term hybridizes the excited states $\ket{+1}$ and $\ket{-1}$, lifting their energy degeneracy and introducing a fitted splitting of $\sim 2.6$ GHz, which manifests as a level anti-crossing in the excited-state ODMR spectrum (Fig. \ref{fig3}(b)). Furthermore, at this anti-crossing point, the $E_1$ term drives mixing between the $\ket{0}$ and $\ket{\pm 1}$ states. The resulting hybridization modifies the effective excited-state lifetimes $\tau_j$, because the $\ket{0}$ and $\ket{\pm 1}$ states possess distinct intrinsic lifetimes (Fig. \ref{fig3}(d), SI.II(c)). The application of an axial magnetic field suppresses this strain-induced state mixing, thereby altering the optical spin initialization dynamics and providing a means to actively control the spin polarization.

Through the integration of our model with experimental data, we have successfully reconstructed the NV center Hamiltonian and characterized its spin-optical dynamics under distinct strain environments. The model exhibits excellent agreement with experimental observations.
By jointly fitting the $B_z$-dependent excited-state energy shifts and lifetimes using the Hamiltonian in equation(\ref{eq1}) (Fig. \ref{fig3} (a, b, d)), we quantitatively extract the $|D|$, $|E_1|$, $|E_2|$ terms (Fig. \ref{fig3} (e)).
Leveraging this strain characterization, we further perform a joint fitting of the multi-pulse polarization dynamics under both zero and high magnetic fields (SI.II(d), Fig. S7). We obtained the underlying transition rates for both symmetry-preserving (Site I) and symmetry-breaking (Site IV) strain environments. All extracted rates are summarized in End Matter Table.\ref{Table1} and correspond to the transitions labeled in End Matter Fig.\ref{fig5}.

 \begin{figure*}[htbp]
  \includegraphics[width=0.9\linewidth]{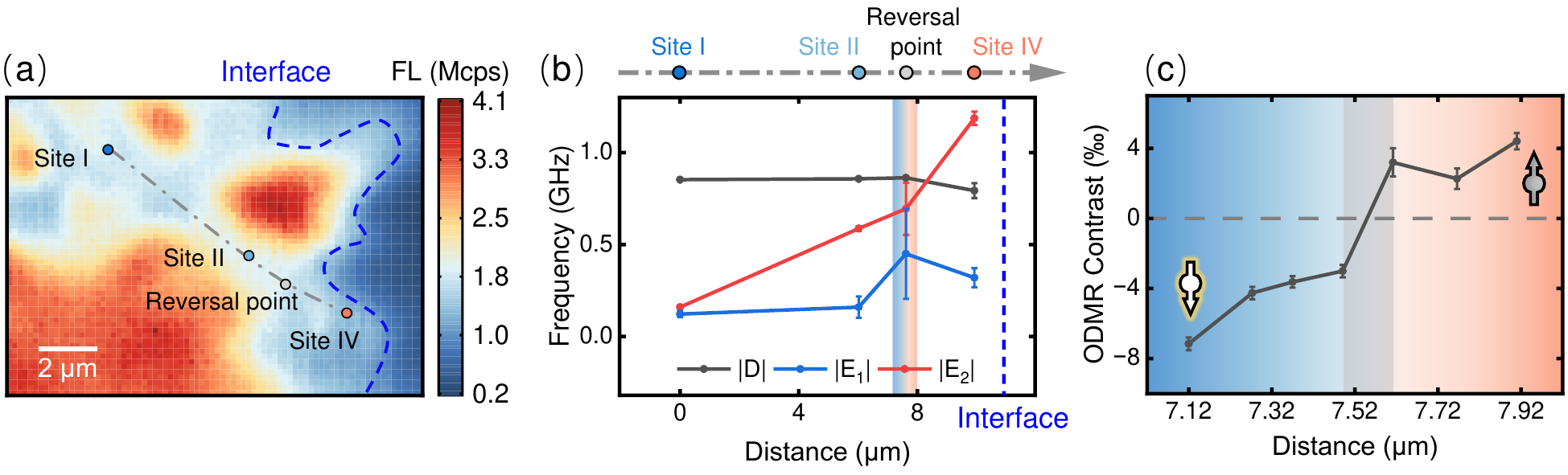}
  \caption{\label{fig4} Nanoscale spin polarization reversal induced by a strain gradient.
  (a) Fluorescence images of NV centers within DAC. A blue dashed line marks the Metal-Powder interface mentioned in Fig.1(b). Sites I, II, IV and an inversion  point are identified along the grey dashed-dotted line from the center to the edge. The reversal point denotes the location where the ODMR contrast transitions from negative to positive. 
  (b) The moduli of $D$, $E_1$ and $E_2$ terms as a fucntion of position. the Metal-Powder interface is marked with a blue dashed line.
  (c) ODMR contrast as a function of distance in the vicinity of the reversal point (corresponding to blue-red gradient regions in (b)). The contrast reverses from negative to positive over a spatial scale of $\sim$ 120 nm (grey rectangle), demonstrating a nanoscale spin polarization reversal driven by the local strain gradient. 
  The error bars in (b) and (c) represent the standard errors of measured values.
  }
  \end{figure*}
 
A comparison of the fitting results from Site I and Site IV reveals the profound impact of symmetry-breaking strain on the spin-optical dynamics of this quantum emitter. Symmetry-breaking strain not only induces spin-state hybridization but also fundamentally reprograms the intersystem crossing (ISC) process, resulting in the spin polarization reversal at Site IV. 
(1) At low magnetic field ($B_z < 100$ G), the significant $|E_1|$ and $|E_2|$ terms at Site IV, induced by strong symmetry-breaking strain, cause substantial mixing among $\ket{0}$ and $\ket{\pm 1}$ excited states. This mixing significantly suppresses the spin selectivity of the up-ISC process. 
(2) The severe deviation of the down-ISC branch ratio $q_0$ to the $\ket{0}$ ground state, which we find to be $q_0 \sim 0.21$ at Site IV---is significantly lower than the typical value of $q_0 \sim 0.54$ in unstrained NV centers\cite{robledo2011spin}. The in-plane distortion-induced symmetry breaking may couple to vibrational modes to mediate the anomalous down-ISC pathway \cite{jin2025first}.
These two conditions lead to net polarization into the dark states $\ket{\pm 1}$  following repeated excitation cycles, thus producing a positive ODMR contrast under resonant microwave driving. At high axial magnetic fields ($B_z > 100 $ G), the first condition is suppressed. The lift of state mixing restores $\ket{0}$-state polarization and the characteristic negative ODMR contrast.

This strain-engineered dynamics control and polarization switching enables a new method for nanoscale quantum emitter modulation, with implications ranging from  programmable on-chip quantum light sources\cite{gao2015coherent,aharonovich2016solid} to high-density spin-based quantum memory\cite{fuchs2011quantum,mouradian2015scalable}. To demonstrate the nanoscale spin polarization reversal under a substantial strain gradient, we analyze NV centers along a radial line from the DAC center to the metal-powder interface (Fig. \ref{fig4}(a)). The ODMR spectra for Sites I, II and IV reveal an reversal point where the ODMR contrast transitions from negative to positive (Fig. \ref{fig1}(d,f,g)). The $|D|$, $|E_1|$, $|E_2|$ terms of each site show distinct spatial dependence (Fig. \ref{fig4}(b)). $|E_2|$ exhibits a sharp increase at ~7.6 $\mu$m, indicating a large symmetry-breaking strain gradient near the metal–powder interface. 
We finally conduct ODMR mapping across the $\pm$ 400 nm region around the reversal point. As depicted in Fig. \ref{fig4}(c), the ODMR contrast switched from positive to negative over the distance range of 7.49 - 7.61 $\mu$m  (marked by the grey rectangle), corresponding to a nanoscale spin polarization reversal. Constrained by the optical resolution limit, we deduce that the reversal occurs on a spatial scale smaller than 120 nm. 

Our results demonstrate that controlled strain engineering enables reversible spin polarization switching in NV centers at the nanoscale. By combining high-pressure techniques with ODMR measurements, we established that symmetry-breaking strain gradients can locally modify ISC dynamics, leading to a complete inversion of the ground-state spin polarization. The observed transition occurs over remarkably short length scales ($\leq$120 nm), enabled by the strong strain gradients near DAC boundaries. This work provides the first direct evidence of strain-mediated spin control at dimensions below the optical diffraction limit, with several important implications: (1) The opposite polarization states at nanoscale suggests exciting possibilities for encoding spin-based information at ultrahigh densities; (2) The strain-polarization coupling mechanism could enable new approaches for quantum sensing of nanoscale strain fields; and (3) The demonstrated reversal dynamics may inform the design of hybrid quantum devices combining strain-tunable spins with photonic or mechanical resonators. The future integration of super-resolution techniques and nanoscale engineering of NV centers could further elucidate the atomistic details of strain-spin coupling, potentially revealing new regimes of spin-stress interactions at the nanoscale.

\textit{Acknowledgment}. This work was supported by the National Natural Science Foundation of China (grants no. T2325023, 92265204, 12474500, 12261160569, 12204485, and 12504570), the Innovation Program for Quantum Science and Technology (grant no. 2021ZD0302200), the Fundamental Research Funds for the Central Universities, and the Postdoctoral Fellowship Program and China Postdoctoral Science Foundation (grant no.BX20240347).

\bibliography{bibfile_formal}

\renewcommand\thefigure{\Alph{section}\arabic{figure}}
\renewcommand\thetable{\Alph{section}\arabic{table}}
\renewcommand{\tablename}{TABLE.}
\clearpage
\newpage
\section{End Matter}
\subsection{Symmetry and strain}
Within the elastic range of diamond, stress and strain exhibit a linear relationship. Here we study the stress tensor to provide an intuitive view of the stress effect on the NV Hamiltonian(\ref{eq1})\cite{udvarhelyi2018spin}. Considering NV center is a point defect of $C_{3v}$ symmetry, we seperate the $3\times3$ stress tensor into two parts: 
\begin{align}
\bm{\sigma} &= \bm{\sigma}_{\rm{C_{3v}}}+\bm{\sigma}_{\rm{non-C_{3v}}}\\
&=\underbrace{\begin{bmatrix}
  \sigma_{m} & 0 & 0 \\
  0 & \sigma_{m} & 0 \\
  0 & 0 & \sigma_{zz}
\end{bmatrix}}_{\rm{symmetry-preserving}}
+
\underbrace{\begin{bmatrix}
  \sigma_{d} & \sigma_{xy} & \sigma_{xz} \\
  \sigma_{yx} & -\sigma_{d} & \sigma_{yz} \\
  \sigma_{zx} & \sigma_{zy} & 0
\end{bmatrix}}_{\rm{symmetry-breaking}}
\end{align},
where $\sigma_{ij}$ represents the stress component in the NV frame (Fig. \ref{fig1}(a)); $\sigma_{m} = (\sigma_{xx}+\sigma_{yy})/2 $ and $\sigma_{d} = (\sigma_{xx}-\sigma_{yy})/2$ represents the in-plane mean normal stress and deviatoric normal stress respectively. \\
(1) The symmetry-preserving stress component  $\sigma_{\rm{C_{3v}}}$ comprises in-plane uniform contraction and axial compress. Under this stress, the degeneracy of the states remains unchanged and the zero-field splitting $D$ undergoes a shift (Fig. \ref{fig1}(h)):
\begin{align}
D = D_0+ g_{41}\sigma_{m} + g_{43}\sigma_{zz}
\end{align},
where $D_0$ denotes the zero-field splitting of NV centers under 
ambient condition.
(2) The symmetry-breaking stress component  $\bm{\sigma}_{\rm{non-C_{3v}}}$ includes xy-plane deviatoric stress and shear components. Under this stress, $E_1$ and $E_2$ couplings in the Hamiltonian are modulated, driving spin-state hybridization and degeneracy lifting:
\begin{align}
E_1 &= (g_{26}\sigma_{zx}-g_{25}\sigma_{d})-\rm{i}(g_{26}\sigma_{yz}+g_{25}\sigma_{xy})\\
E_2 &= (g_{16}\sigma_{zx}-g_{15}\sigma_{d})-\rm{i}(g_{16}\sigma_{yz}+g_{15}\sigma_{xy})
\end{align}
where i denotes imaginary unit and $g_{ij}$ represents spin-stress coupling-strength parameters.

\subsection{Spin Dynamics fitting} For both symmetry-preserving strain (Site I) and symmetry-breaking strain (Site IV) configurations, key parameter---including excited-state lifetimes, the moduli of $D$, $E_1$ and $E_2$ terms in the excited-state Hamiltonian and spin dynamics fitting results---are summarized in End Matter Table. \ref{Table1}. A combined approach that integrates ODMR spectroscopy and time-resolved fluorescence dynamics measurements facilitates the establishment of  spin-dynamic models, as well as their coupling with strain and magnetic field. Three models are included in this workflow: 

(1) Tri-exponential decay models are used to describe the fluorescence decay of NV center induced by ps-laser pulse (SI.II(b)). Following deterministic spin-state preparation and manipulation, the fluorescence decay measurements yield the lifetimes $\tau_j$ (Fig. \ref{fig3}(d)) and the excited-state populations $P_j$ (Fig. \ref{fig2}). 

(2) The excited-state NV Hamiltonian under strain and magnetic fields governs the energy levels structure and the corresponding eigenstates, thereby establishing a quantitative relationship between environmental parameters ($B_z$ and strain) and experimental observables, including ODMR spectra and excited state lifetimes (SI.II(c)). Supplemented by the linear relationship between excited-state mixing and lifetimes, this framework enables quantitative determination of magnitudes of  the magnitudes of $|D|$, $|E_1|$, $|E_2|$ terms in the excited-state Hamiltonian (Fig. \ref{fig3}).

\begin{figure*}
  \includegraphics[width=0.3\linewidth]{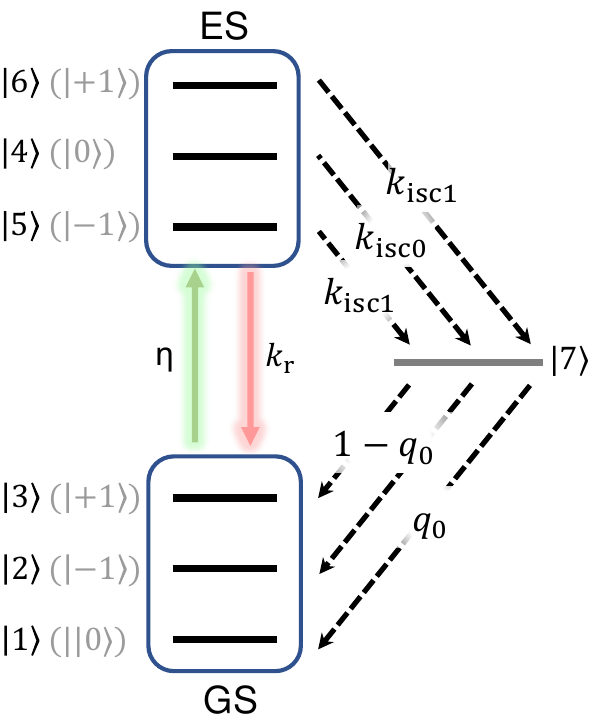}
  \caption{\label{fig5} The energy level structure of NV centers under high magnetic field ($B_z >700$ G). The spin states are purified into $\ket{0}$, $\ket{\pm 1}$ under high $B_z$. The green arrows denote the 532-nm laser-induced excitation from ground states to excited states. The red arrows represent the spontaneous emission Process. The dashed-line arrow represents the spin-dependent ISC process.
     $\eta$: excitation probability.
     $k_r$: spontaneous emission rates.
     $k_{\rm{isc0}}$, $k_{\rm{isc1}}$: up-ISC transition rates from pure excited state $\ket{0}$, $\ket{\pm 1}$ to the metastable singlet state $\ket{7}$.
     $q_0$: down-ISC transition probability from singlet state to ground state $\ket{0}$.
  }
  \end{figure*}

(3) The multi-pulse polarization dynamics recursive model provides detailed spin dynamics information (SI.II(d)). By monitoring the population evolution through a multi-pulse sequence enables the quantitative extraction of spin dynamics parameters ($k_r$, $k_{\rm{isc0}}$, $k_{\rm{isc1}}$, $q_0$). The parameters and the corresponding transition are marked in End Matter Fig. \ref{fig5}. Fitting results are provided in End Matter Table. \ref{Table1}.

\newcommand{\tabincell}[2]{\begin{tabular}{@{}#1@{}}#2\end{tabular}}  
\begin{table*}
\centering
\begin{ruledtabular}
\begin{tabular}{cllll|l}
        Measurements            & Parameters                          &                   & Site I           & Site IV          & Amb. NV \\
        \colrule
        
        \multirow{2}{*}{\tabincell{c}{Fluorescence decay\\measurement}}&\multirow{2}{*}{\tabincell{l}{Excited-state\\lifetimes}} 
                                                        & $\tau_{\rm{bright}}$ (ns) &6.12\ $\pm\ 0.02$ &6.59\ $\pm\ 0.03$ & 13.7    \\
                                                        & & $\tau_{\rm{dark}}$ (ns)   &2.05\ $\pm\ 0.01$ &2.32\ $\pm\ 0.03$ & 8.6    \\
                                                        \\
        
        \multirow{3}{*}{\tabincell{c}{ODMR-lifetimes\\joint fitting}}&\multirow{3}{*}{\tabincell{l}{Strain\\terms in\\ Excited-State $\hat H$}}   & $|D|$ (GHz)     &0.85\ $\pm\ 0.01$ &0.80\ $\pm\ 0.06$ & 1.4     \\
                                                        & & $|E_1|$ (GHz)   &0.09\ $\pm\ 0.01$ &0.25\ $\pm\ 0.04$ &  0       \\
                                                        & & $|E_2|$ (GHz)   &0.16\ $\pm\ 0.02$ &1.19\ $\pm\ 0.03$ &  0     \\
                                                        \\                   
        \multirow{4}{*}{\tabincell{c}{Multi-pulse\\polarization\\ measurement}}&\multirow{4}{*}{\tabincell{l}{Dynamic\\parameters}}  & $k_r$ (MHz)     &132\ $\pm\ 2$     &150\ $\pm\ 1$     & 67.7\ $\pm\  3.4$\\
                                                        & & $k_{\rm{isc0}}$ (MHz)&32\ $\pm\ 2$      &2 $\pm\ 1$        &6.4\ $\pm\ 2.3$\\
                                                        & & $k_{\rm{isc1}}$ (MHz)&357\ $\pm\ 3$     &282 $\pm\ 6$      &50.7\ $\pm\ 4.4$\\
                                                        & & $q_0$           &0.39\ $\pm\ 0.01$ &0.21\ $\pm\ 0.01$ &0.54\ $\pm\ 0.22$
                                                        \\
                                                        
\end{tabular}
\caption{Summary of the fitting result of the spin-optical dynamics transition rates. The measurements are listed in the order of workflow. $\tau_{\rm{bright}}$ is defined as the lifetimes of the bright state $\ket{0}$, while $\tau_{\rm{dark}}$ represents the mean lifetime of the dark state $\ket{\pm 1}$. Under high magnetic field, $\tau_{\rm{bright}} = \tau_4$, $\tau_{\rm{dark}}= \frac{1}{2}(\tau_5+ \tau_6)$. $|D|$, $|E_1|$, $|E_2|$ denote the moduli of the spin-strain coupling terms in the excited-state Hamiltonian. $k_r$, $k_{\rm{isc0}}$, $k_{\rm{isc1}}$, $q_0$ are the spin dynamics parameters marked in End Matter Fig. \ref{fig5}. The parameters of the NV centers under ambient conditions \cite{robledo2011spin} are shown for comparison.}
\label{Table1}
\end{ruledtabular}
\end{table*}

\end{document}